\documentclass[letterpaper]{article} 
\usepackage{aaai2026}  
\usepackage{times}  
\usepackage{helvet}  
\usepackage{courier}  
\usepackage[hyphens]{url}  
\usepackage{graphicx} 
\usepackage{natbib}  
\usepackage{caption} 
\usepackage{algorithm}
\usepackage{algorithmic}

\usepackage{newfloat}
\usepackage{listings}
\DeclareCaptionStyle{ruled}{labelfont=normalfont,labelsep=colon,strut=off} 
\floatstyle{ruled}
\newfloat{listing}{tb}{lst}{}
\floatname{listing}{Listing}
\title{Uncovering Latent Connections in Indigenous Heritage:\\Semantic Pipelines for Cultural Preservation in Brazil}
\author {
    Luis Vitor Zerkowski\textsuperscript{\rm 1},
    Nina S. T. Hirata\textsuperscript{\rm 2}
}
\affiliations {
    \textsuperscript{\rm 1}University of Amsterdam\\
    \textsuperscript{\rm 2}University of São Paulo\\
    luisvz@gmail.com, nina@ime.usp.br
}

\begin{document}

\maketitle

\begin{abstract}

Indigenous communities face ongoing challenges in preserving their cultural heritage, particularly in the face of systemic marginalization and urban development. In Brazil, the \emph{Museu Nacional dos Povos Indígenas} through the \emph{Tainacan} platform hosts the country’s largest online collection of Indigenous objects and iconographies, providing a critical resource for cultural engagement. Using publicly available data from this repository, we present a data-driven initiative that applies artificial intelligence to enhance accessibility, interpretation, and exploration. We develop two semantic pipelines: a visual pipeline that models image-based similarity and a textual pipeline that captures semantic relationships from item descriptions. These embedding spaces are projected into two dimensions and integrated into an interactive visualization tool we also developed. In addition to similarity-based navigation, users can explore the collection through temporal and geographic lenses, enabling both semantic and contextualized perspectives. The system supports curatorial tasks, aids public engagement, and reveals latent connections within the collection. This work demonstrates how AI can ethically contribute to cultural preservation practices.

\end{abstract}

\begin{links}
    \link{Code}{https://github.com/Luizerko/indigenous_clusters_and_communities}
    \link{Data}{https://tainacan.museudoindio.gov.br/}
\end{links}

\section{Introduction}

Indigenous peoples in Brazil continue to face cultural erasure, land dispossession, and limited visibility despite constitutional protections and growing public awareness. Preserving Indigenous knowledge - its languages, symbols, and histories - is both an ethical imperative and a technological challenge. Public institutions have begun to digitize important collections to support cultural preservation; among them, the \emph{Museu Nacional dos Povos Indígenas} \cite{museudoindio2025} through the \textit{Tainacan} platform \cite{tainacan2025} hosts Brazil’s largest online repository of Indigenous objects and iconographies. While this represents a major step forward, the collection remains difficult to explore due to limited semantic organization and interactivity.

This work proposes a data-driven system that enhances access to this digital collection using artificial intelligence. We present two semantic pipelines: one that models visual similarity using features extracted from item images, and another that captures semantic relationships from textual descriptions. Each pipeline produces an embedding space, projected to two dimensions for interactive visualization. These structured representations are integrated into a developed point-cloud interface that enables users to explore the collection through similarity-based navigation.

In addition to semantic exploration, the tool offers browsing through temporal and geographic lenses, offering more traditional, contextualized views of the data. The system supports both public engagement and curatorial analysis by revealing mislabeled items, surfacing latent connections, and suggesting new interpretive directions. It is fully open-source and designed for transparency and reuse, with code, documentation, and dataset references available at \url{will_be_available_on_camera-ready}.

\subsection{Related Work}

Preserving Indigenous heritage is both a cultural and technological challenge, particularly in countries like Brazil with immense Indigenous diversity. Digital technologies have played a growing role in this space, enabling broader accessibility and engagement \cite{munoz2020digital}. In education, \textit{Guimarães} \cite{guimaraes2015teaching} and \textit{Ferreira} \cite{ferreira2011brazilian} emphasize the need to integrate Indigenous perspectives dynamically rather than as static representations, while others highlight connections to biodiversity and linguistic preservation \cite{wilder2016importance}.

Platforms like \emph{Mukurtu CMS} \cite{mukurtu2025}\footnote{\url{https://mukurtu.org/}} and repatriation projects \cite{rebellato2025digital} aim to involve Indigenous communities in curatorial processes. Still, most remain reliant on metadata-driven interfaces, which can limit deeper semantic engagement \cite{almaaitah2020opportunities}. To address this, researchers have explored semantic modeling and AI-based design to enrich representation fidelity and interactivity.

Recent progress in computer vision and language modeling offers a technical foundation for working with cultural data. Vision models such as \texttt{U-Net} \cite{ronneberger2015unet} and modern transformers like \texttt{ViT} \cite{dosovitskiy2021an}, \texttt{SAM} \cite{kirillov2023segment}, and \texttt{DINOv2} \cite{oquab2023dinov2} enhance image-based analysis. In the textual domain, models like \texttt{BERTimbau} \cite{souza2020bertimbau} and \texttt{Albertina} \cite{rodrigues2023advancing}, often trained with contrastive learning \cite{gao-etal-2021-simcse, liao2021sentence}, enable nuanced sentence-level embeddings. Cross-modal models such as \texttt{CLIP} \cite{radford2021learning} open up further opportunities for aligning imagery and text.

To make these embeddings interpretable and user-friendly, dimensionality reduction techniques like \texttt{UMAP} \cite{McInnes2018}, \texttt{t-SNE} \cite{vanDerMaaten2008}, and \texttt{TriMap} \cite{amid2019trimap} are often used to project embedding spaces into 2D or 3D plots. Tools such as the Embedding Projector \footnote{\url{https://projector.tensorflow.org/}} \cite{smilkov2016embedding} exemplify how interactive visualizations can enhance exploration. Similarly, interpretability-focused initiatives like the Activation Atlas \footnote{\url{https://distill.pub/2019/activation-atlas/}} \cite{carter2019activation} offer engaging, exploratory views of complex models and datasets.

Importantly, Indigenous digital protagonism \cite{hisayasu2019mediated} stresses the active role of Indigenous communities in shaping how their histories are represented online, reinforcing the need for tools that are not only technically robust, but also culturally inclusive by design.

\subsection{Contributions}

This work offers three main contributions. First, we developed a complete data processing pipeline for the Indigenous collection of the \emph{Museu Nacional dos Povos Indígenas}, including documentation, practical guidelines, and machine-learning-ready datasets with background-removed images and summarized descriptions, all shared with the museum. Second, we introduced a modular dual-pipeline framework for modeling visual and textual similarity, adapted to Brazilian Portuguese and the visual specificity of Indigenous items. The framework supports various architectures and training strategies and is designed for reuse across collections. Lastly, we implemented an interactive semantic visualization tool for exploring the collection through embedding point-clouds, as well as temporal and geographic views. Designed for both public and curatorial use, it is scheduled for deployment on Brazil’s federal government website.

\section{Data Gathering and Processing}
\label{sec:data}

The dataset employed in this work was extracted from the public \emph{Tainacan} repository \cite{tainacan2025}, with a focus on the collection curated by the \emph{Museu Nacional dos Povos Indígenas} \cite{museudoindio2025}. Metadata from 20,965 items was collected via custom \texttt{Bash} and \texttt{Python} scripts, with data usage authorized by the museum. Key attributes include the item's category (\texttt{categoria}), Indigenous community (\texttt{povo}), thumbnail image (\texttt{thumbnail}), and full-text description (\texttt{descricao}).

Following extraction, the dataset underwent preprocessing tailored to visual and textual modeling pipelines. For images, we applied background removal using the \texttt{RMBG-2.0} model \cite{BiRefNet}. This step mitigated risks of spurious visual correlations across the 11,274 items (the ones with image data). Text descriptions were lowercased and then, to meet \texttt{GPU} memory constraints (\texttt{8GB VRAM}) and enhance representation learning, we implemented a summarization pipeline targeting a 64-token maximum. This was achieved using the \texttt{Llama-4-Maverick-17B-128E-Instruct} model \cite{touvron2023llama}, accessed via the \texttt{Groq API} \cite{groq2025} \footnote{\url{https://groq.com/}}. The prompts emphasized semantic retention, prioritizing details and cultural vocabulary.

Building on the summarized corpus, we constructed contrastive learning datasets. For the supervised setup, each anchor was paired with a paraphrased positive, generated via the aforementioned large language model (LLM), and ten negative examples randomly sampled from different \texttt{categoria} labels, to ensure semantic dissimilarity.

Now for evaluating both supervised and unsupervised models, we also developed a benchmark called \texttt{In-Context STS-B}. For each item, we used the LLM-generated paraphrase as a positive example, but additionally asked the LLM to select the most semantically distinct description of a chunk as a negative example. Because our chunks were originally designed for description summarization though, they contained sequential data. While this occasionally led to highly similar descriptions grouped and hence some false negatives, the dataset still offered useful insights into contextualized semantic sensitivity.

\section{Constructing Semantic Spaces}

With the dataset fully processed, we turn to constructing semantic spaces that reflect the visual and textual similarities among cultural artifacts. Our goal is twofold: to build practical tools that facilitate intuitive exploration of large archival datasets, and to demonstrate how machine learning can augment human curation by revealing latent structures within collections. This section begins with baselines grounded in domain knowledge. These methods do not rely on learning from data but instead leverage metadata and interpretable encodings to establish reference points. While limited in representational capacity, they were thought of to validate the output of more complex models, but also illustrate the limitations of purely categorical representations.

\subsection{Baselines from Metadata Encodings}

Before arriving at the final pipeline, we tested several metadata-based strategies to explore direct semantic relationships within the collection. These included orthonormal projections of categorical attributes, \texttt{K-Modes} clustering \cite{huang1998extensions} over multi-hot metadata, and even a triangular layout based solely on \texttt{tipo\_de\_materia\_prima}. While each method was analytically interpretable, none yielded sufficiently coherent or generalizable semantic groupings. Dimensionality reduction techniques such as \texttt{UMAP} \cite{McInnes2018}, \texttt{TriMap} \cite{amid2019trimap}, and \texttt{t-SNE} \cite{vanDerMaaten2008} also produced noisy, meaningless projections. We further explored the use of language family trees as a reference taxonomy but found them too fragmented and inconsistent for computational use.

This reframed the role of our semantic models: rather than validating against fixed baselines, they become tools for surfacing latent patterns and generating curatorial hypotheses. Both visual and textual embeddings support the discovery of connections and inconsistencies not captured by existing metadata. In this way, our platform offers a starting point for deeper curatorial inquiry and interdisciplinary collaboration. Full implementation details and evaluations of the early strategies can be found in the extended version of this work and in the public repository documentation.

\subsection{Image-Based Semantic Spaces}

To understand visual relationships among collection items, we developed a deep learning-based image pipeline. We extracted embeddings from background-removed images using two transformer-based architectures, \texttt{ViT Base} \cite{dosovitskiy2021an} and \texttt{DINOv2 Base} \cite{oquab2023dinov2}, and projected the resulting representations into 2D for qualitative analysis. Among dimensionality reduction techniques tested, \texttt{UMAP} yielded the clearest manifolds.

\paragraph{Fine-Tuning Strategy:}  
Our initial approach to fine-tuning (a second one will be introduced later for textual embeddings) relied on simple supervised learning using existing attributes in the dataset as targets. While the pretrained visual models already displayed reasonable semantic structure, verified through qualitative inspection, we sought to further refine this structure and inject higher-level semantic alignment. To achieve this, we used supervised training with the \texttt{povo} and \texttt{categoria} attributes as targets.

We followed standard fine-tuning practices from the \texttt{ViT} literature \cite{dosovitskiy2021an, steiner2022train}, attaching a single linear classification head on top of the transformer backbone. The same architecture was used for \texttt{DINOv2} to enable fair comparison. Models were trained until convergence, typically between 10 and 20 epochs, with early stopping applied. While we explored a broad range of hyperparameters, we report only the most representative and effective configurations - those sufficient to illustrate performance trends across different values. Each configuration was trained multiple times to report mean and standard deviation across runs.

Class imbalance was a major challenge, especially for \texttt{povo}, which includes 187 classes, many with only a handful of samples. To address this, we implemented a rebalancing strategy combining class filtering, data augmentation, and weighted loss functions, a combination that consistently improved performance across evaluation metrics. We also experimented with layer freezing (starting from the bottom up) to preserve general knowledge in early layers.

We later introduced a multi-head setup to jointly optimize for both \texttt{povo} and \texttt{categoria}, aiming to explore whether combining distinct semantic signals would enrich the learned representations. For this setup, all training used a rebalanced dataset, focusing solely on \texttt{povo}, since attempting to rebalance both attributes either reintroduced imbalance in the other or failed due to extreme sparsity in the joint distribution. As discussed later, this configuration yielded particularly interesting numerical and qualitative results.

\paragraph{Evaluation:}  
Since the original models lacked classification heads, we couldn't numerically compare performances with them. For quantitative analysis on the fine-tuned models, however, we measured accuracy, precision, and recall across all trained models. Given the severe class imbalance, particularly for \texttt{povo}, we introduced refined metrics - Precision/Recall (Selected) - which evaluate performance only on the subset of classes used in the rebalanced dataset. This allows a fair comparison between models trained with and without rebalancing, as evaluating on unseen classes would be misleading for the rebalanced models.

Finally, we visually examined the embedding spaces produced by each model to assess not only dispersion but also the emergence of coherent clusters aligned with the target attributes. This qualitative inspection served as an additional layer of validation for the models’ representational quality.

\paragraph{ViT-Based Models:} Results are summarized in Tables \ref{tab:vit-povo}, \ref{tab:vit-categoria} and \ref{tab:vit-multihead}.

\begin{table}[h]
\centering
\scriptsize
\begin{tabular}{c c c c c}
\cline{1-5}
\textbf{Training Data} & \textbf{Frozen (\%)} & \textbf{Acc. (\%)} & \textbf{Prec. Sel.} & \textbf{Rec. Sel.} \\
\cline{1-5}
Original & 0 & \underline{68.32 ± 1.52} & 0.58 ± 0.01 & \underline{0.62 ± 0.01} \\
Rebalanced & 0 & \textbf{70.99 ± 0.73} & \textbf{0.71 ± 0.02} & \textbf{0.69 ± 0.02} \\
Rebalanced & 80 & 67.48 ± 1.13 & 0.64 ± 0.01 & \underline{0.62 ± 0.01} \\
\end{tabular}
\caption{\small Test set performance of fine-tuned \texttt{ViT} models on the \texttt{povo} classification task. Models were (mostly) trained with a learning rate of \(2 \times 10^{-5}\), a weight decay of \(2 \times 10^{-6}\), and weighted loss inversely proportional to class frequency.}
\label{tab:vit-povo}
\end{table}

\begin{table}[h]
\centering
\scriptsize
\begin{tabular}{c c c c c}
\cline{1-5}
\textbf{Training Data} & \textbf{Frozen (\%)} & \textbf{Acc. (\%)} & \textbf{Prec. Sel.} & \textbf{Rec. Sel.} \\
\cline{1-5}
Original & 0 & \underline{87.60 ± 0.81} & \underline{0.86 ± 0.02} & \underline{0.84 ± 0.01} \\
Rebalanced & 0 & \textbf{88.64 ± 1.53} & \textbf{0.88 ± 0.01} & \textbf{0.85 ± 0.01} \\
Rebalanced & 80 & 86.30 ± 1.30 & 0.84 ± 0.02 & 0.83 ± 0.02 \\
\end{tabular}
\caption{\small Test set performance of fine-tuned \texttt{ViT} models on the \texttt{categoria} classification task. Models were (mostly) trained with a learning rate of \(3 \times 10^{-6}\), a weight decay of \(1 \times 10^{-6}\), and weighted loss inversely proportional to class frequency.}
\label{tab:vit-categoria}
\end{table}

\begin{table}[h]
\centering
\scriptsize
\begin{tabular}{c c c c}
\cline{1-4}
\textbf{Head Wgts.} & \textbf{Acc. (\%)} & \textbf{Prec. Sel.} & \textbf{Rec. Sel.} \\
\cline{1-4}
50 / 50 & \textbf{71.38} / 88.16 & \textbf{0.72} / \underline{0.86} & \textbf{0.68} / 0.85 \\
70 / 30 & \underline{71.16} / \underline{89.26} & \textbf{0.72} / \textbf{0.89} & \textbf{0.68} / \textbf{0.88} \\
30 / 70 & 70.28 / \textbf{89.40} & \underline{0.69} / \textbf{0.89} & \textbf{0.68} / \underline{0.87} \\
\end{tabular}
\caption{\small Test set performance of fine-tuned \texttt{ViT} models on both \texttt{povo} and \texttt{categoria} classification tasks. For each column, the first result is directed to the \texttt{povo} head and the second one directed to the \texttt{categoria} head. Models were trained with a learning rate of \(1 \times 10^{-5}\), a weight decay of \(3 \times 10^{-6}\) and weighted loss inversely proportional to class frequency for each head.}
\label{tab:vit-multihead}
\end{table}

Rebalancing significantly improved performance, especially in precision and recall for selected classes. In contrast, partially freezing the network reduced performance, suggesting that \texttt{ViT}’s lower layers required adaptation to our dataset.

\begin{figure}[h]
    \centering
    \includegraphics[width=0.22\textwidth]{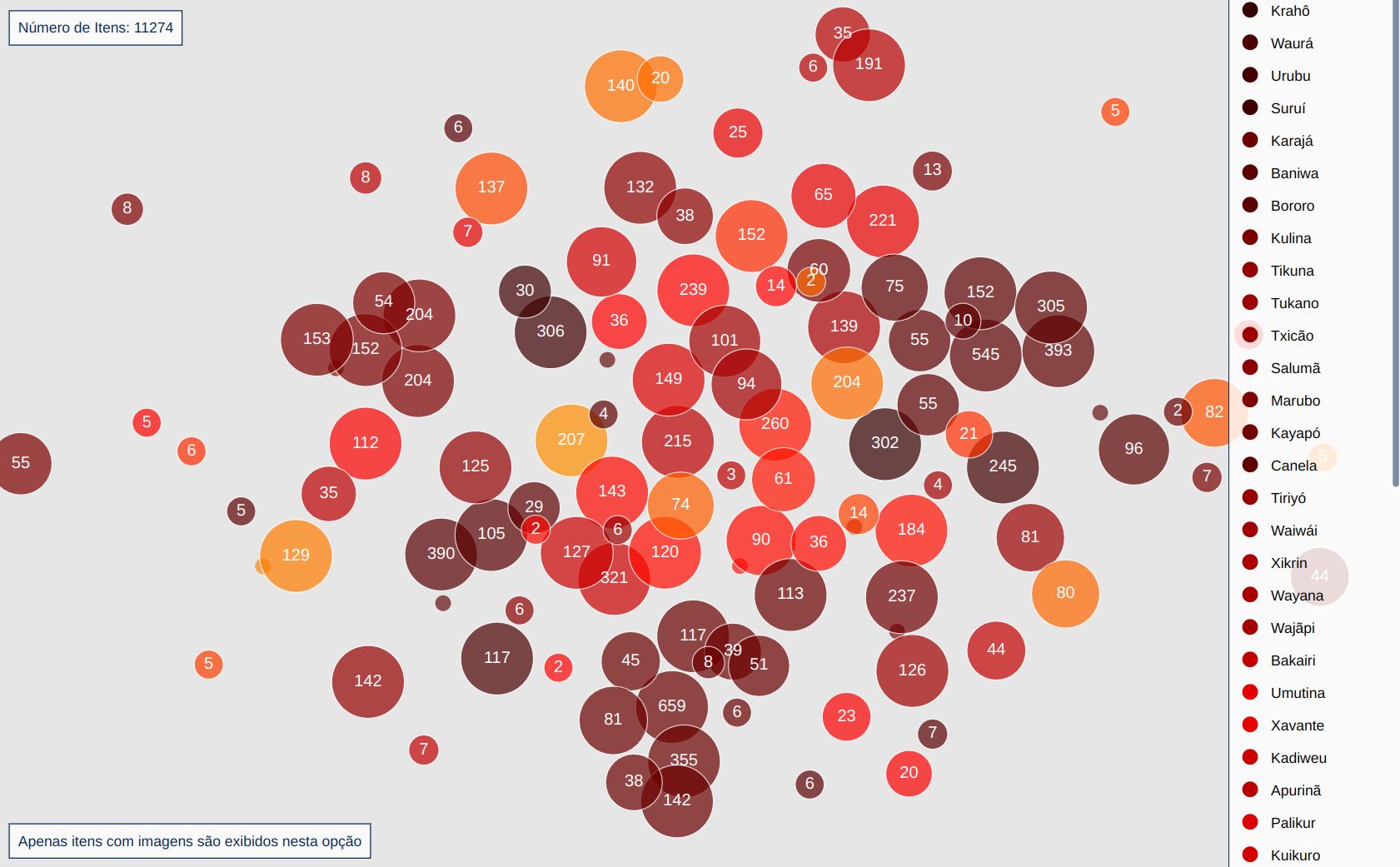}
    \includegraphics[width=0.22\textwidth]{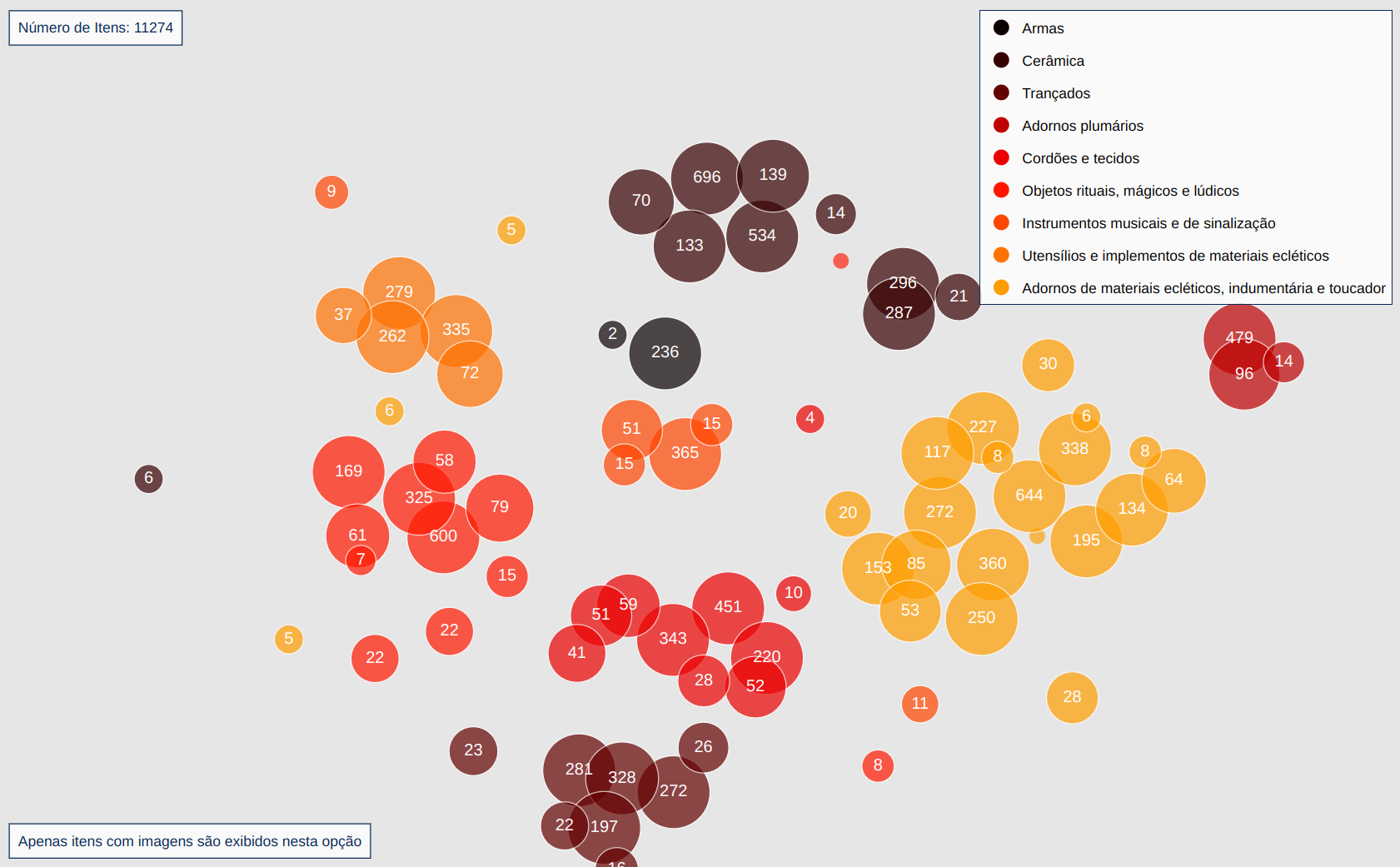}
    \caption{\small Embedding spaces generated by the \texttt{ViT} model fine-tuned on \texttt{povo} (left) and \texttt{categoria} (right). The \texttt{povo} projection shows limited global structure due to the huge amount of classes and class imbalance. The \texttt{categoria} projection displays well-defined clusters and smoother transitions, indicating a coherent underlying manifold.}
    \label{fig:point-clouds-vit}
\end{figure}

Embedding spaces also revealed meaningful structure. While \texttt{povo} lacked global organization due to excluded low-sample classes, local clusters formed for well-represented groups (Figure \ref{fig:point-clouds-vit} left). In the \texttt{categoria} space, coherent groupings and smooth transitions indicated a strong semantic manifold. Even excluded classes like \texttt{etnobotânica} were placed in visually appropriate regions, highlighting the model's generalization ability (Figure \ref{fig:point-clouds-vit} right).

Multi-head training, in turn, generally provided numeric gains, interestingly showing that joint knowledge representation enhanced learning. Single-head models, however, achieved cleaner clusters, revealing more meaningful boundaries and transitions in the embedding space.

\paragraph{DINOv2-Based Models:} Results are summarized in Tables \ref{tab:dino-povo}, \ref{tab:dino-categoria} and \ref{tab:dino-multihead}.

\begin{table}[h]
\centering
\scriptsize
\begin{tabular}{c c c c c}
\cline{1-5}
\textbf{Training Data} & \textbf{Frozen (\%)} & \textbf{Acc. (\%)} & \textbf{Prec. Sel.} & \textbf{Rec. Sel.} \\
\cline{1-5}
Original & 0 & 62.73 ± 0.62 & 0.53 ± 0.02 & 0.65 ± 0.01 \\
Rebalanced & 0 & \underline{71.46 ± 1.13} & \underline{0.70 ± 0.01} & \underline{0.67 ± 0.01} \\
Rebalanced & 80 & \textbf{74.70 ± 1.24} & \textbf{0.74 ± 0.03} & \textbf{0.71 ± 0.01} \\
\end{tabular}
\caption{\small Test set performance of fine-tuned \texttt{DINOv2} models on the \texttt{povo} classification task. Models were (mostly) trained with a learning rate of \(1 \times 10^{-6}\), a weight decay of \(3 \times 10^{-7}\), and weighted loss inversely proportional to class frequency.}
\label{tab:dino-povo}
\end{table}

\begin{table}[h]
\centering
\scriptsize
\begin{tabular}{c c c c c}
\cline{1-5}
\textbf{Training Data} & \textbf{Frozen (\%)} & \textbf{Acc. (\%)} & \textbf{Prec. Sel.} & \textbf{Rec. Sel.} \\
\cline{1-5}
Original & 0 & 86.60 ± 1.44 & \underline{0.84 ± 0.03} & \underline{0.83 ± 0.02} \\
Rebalanced & 0 & \underline{90.21 ± 1.22} & \textbf{0.87 ± 0.02} & \textbf{0.88 ± 0.02} \\
Rebalanced & 80 & \textbf{90.52 ± 0.64} & \textbf{0.87 ± 0.01} & \textbf{0.88 ± 0.01} \\
\end{tabular}
\caption{\small Test set performance of fine-tuned \texttt{DINOv2} models on the \texttt{categoria} classification task. Models were (mostly) trained with a learning rate of \(1 \times 10^{-6}\), a weight decay of \(3 \times 10^{-7}\), and weighted loss inversely proportional to class frequency.}
\label{tab:dino-categoria}
\end{table}

\begin{table}[h]
\centering
\scriptsize
\begin{tabular}{c c c c}
\cline{1-4}
\textbf{Head Wgts.} & \textbf{Acc. (\%)} & \textbf{Prec. Sel.} & \textbf{Rec. Sel.} \\
\cline{1-4}
50 / 50 & \underline{69.61} / \textbf{89.61} & \textbf{0.68} / \textbf{0.87} & \underline{0.66} / \textbf{0.87} \\
70 / 30 & 68.91 / 87.69 & \textbf{0.68} / \underline{0.85} & \underline{0.66} / 0.85 \\
30 / 70 & \textbf{70.86} / \underline{89.05} & \textbf{0.68} / \textbf{0.87} & \textbf{0.67} / \underline{0.86} \\
\end{tabular}
\caption{\small Test set performance of fine-tuned \texttt{DINOv2} models on both \texttt{povo} and \texttt{categoria} classification tasks. For each column, the first result is directed to the \texttt{povo} head and the second one directed to the \texttt{categoria} head. Models were trained with a learning rate of \(3 \times 10^{-7}\), a weight decay of \(1 \times 10^{-7}\) and weighted loss inversely proportional to class frequency for each head.}
\label{tab:dino-multihead}
\end{table}

Even more than the previous experiments, dataset balancing significantly improved performance. In its best configurations, \texttt{DINOv2} consistently outperformed \texttt{ViT}, achieving up to 4\% higher accuracy while maintaining competitive precision and recall. Notably, this time freezing up to 80\% of the model's layers yielded performance gains, suggesting that its pretrained representations are more transferable than those of \texttt{ViT}. Lastly we observe that this model did not numerically benefited from the multi-head setting.

\begin{figure}[h]
    \centering
    \includegraphics[width=0.35\textwidth]{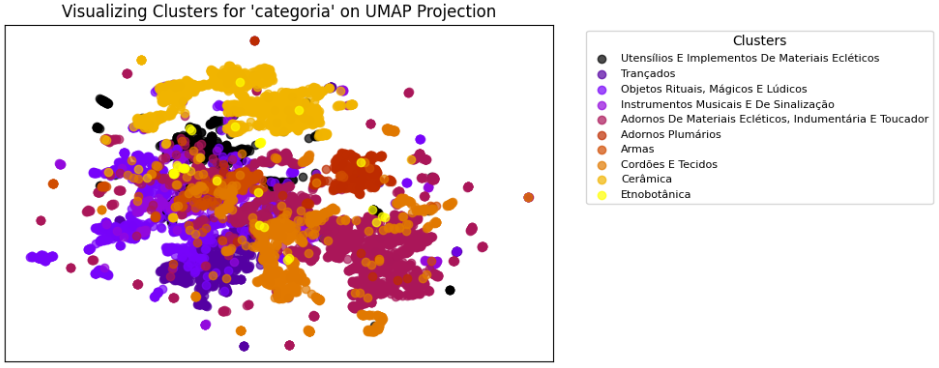}
    \caption{\small Embedding space of the \texttt{DINOv2} multi-head model, colored by \texttt{categoria}. Despite joint supervision, \texttt{categoria} clusters remain interpretable.}
    \label{fig:dino-multihead-categoria}
\end{figure}

Despite the single-head quantitative improvements, \texttt{DINOv2} embeddings were occasionally less interpretable when visualized in two dimensions. However, the multi-head variant preserved more semantic structure, with distinct \texttt{categoria} clusters remaining somewhat visible even under joint optimization (Figure \ref{fig:dino-multihead-categoria}).

\paragraph{Latent Observations:} 

\begin{figure}[h]
    \centering
    \includegraphics[width=0.12\textwidth]{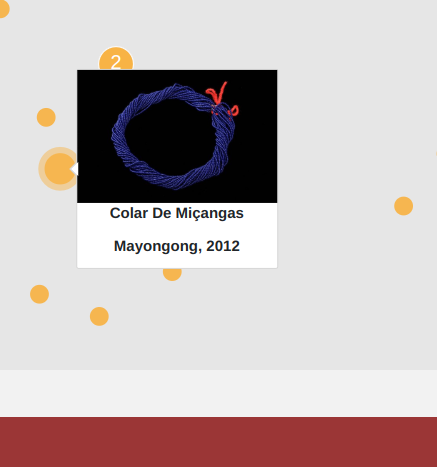}
    \includegraphics[width=0.12\textwidth]{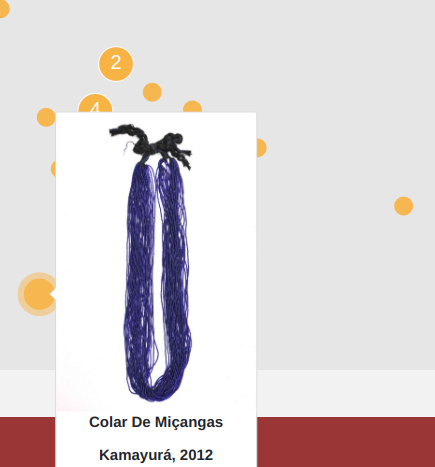}
    \includegraphics[width=0.12\textwidth]{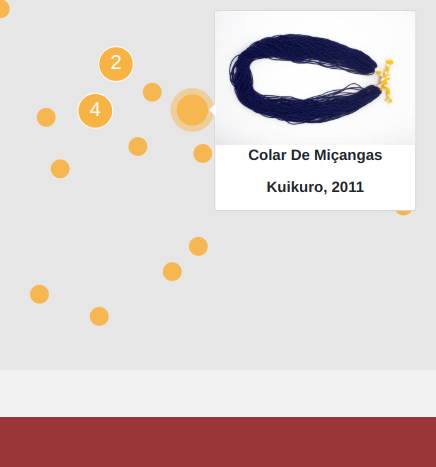}
    \caption{\small Visual similarities between necklaces from the \textit{Mayongong} (left), \textit{Kamayurá} (center), and \textit{Kuikuro} (right) communities.}
    \label{fig:3-necklaces}
\end{figure}

\begin{figure}[h]
    \centering
    \includegraphics[width=0.18\textwidth]{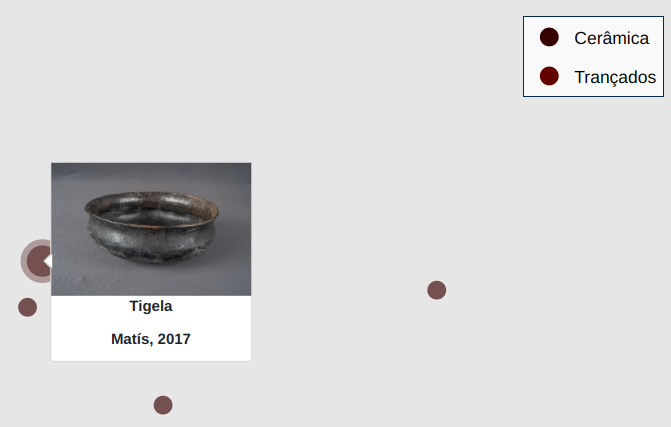}
    \includegraphics[width=0.18\textwidth]{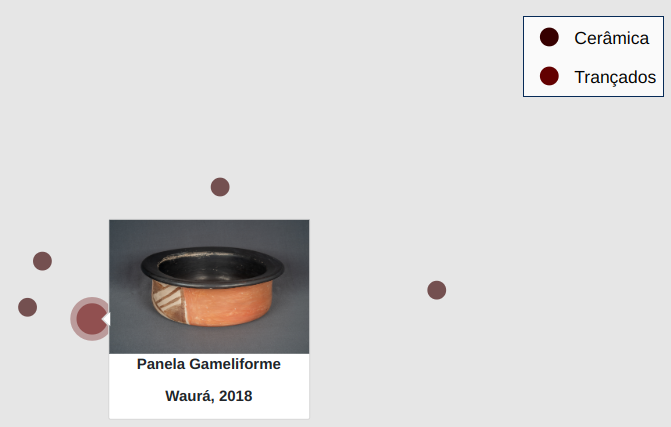}
    \caption{\small Typical ceramic item (left) and a misclassified outlier (right), embedded correctly by the model despite its incorrect \texttt{trançado} label.}
    \label{fig:outlier}
\end{figure}

We highlight a few latent patterns surfaced by a \texttt{ViT} model fine-tuned on \texttt{categoria}. Embeddings revealed that some necklaces from the \textit{Mayongong} community featured vibrant blue palettes, closely aligned with those of \textit{Kamayurá} and \textit{Kuikuro}, suggesting shared material influences (Figure~\ref{fig:3-necklaces}). Outlier detection also exposed mislabeled items, such as a ceramic piece incorrectly tagged as \texttt{trançado}, yet embedded correctly among ceramics (Figure~\ref{fig:outlier}). These cases illustrate how semantic spaces can support curatorial analysis by uncovering hidden relationships and inconsistencies.

\subsection{Text-Based Semantic Spaces}

Building on the image-based framework, we developed a parallel pipeline for textual data using summarized item descriptions. The goal remains consistent: to support semantic navigation and reveal latent relationships within the collection. To generate high-dimensional embeddings, we employed two sentence encoders: \texttt{BERTimbau base} \cite{souza2020bertimbau}, a strong starting point given its prior success in language modeling tasks, and \texttt{Albertina 100M} \cite{rodrigues2023advancing}, a more recent encoder with better Semantic Textual Similarity (\texttt{STS}) \cite{agirre-etal-2012-semeval} scores. We then projected embeddings into 2D using \texttt{UMAP} for analysis. Despite using compact versions of the models due to hardware limitations, models produced coherent and interpretable visual structures.

\paragraph{Fine-Tuning Strategies:} To support richer exploration rather than pure classification, we shifted focus to contrastive learning on encoded sentences for directly modeling semantic similarity, particularly important for textual data, where Indigenous vocabulary challenged pre-trained models. For the unsupervised setup, we followed the unsupervised \texttt{SimCSE} framework \cite{gao-etal-2021-simcse}, using a single dropout layer on top of the models and the \texttt{NT-Xent} loss \cite{chen2020simple}. By running each input through the encoder twice with stochastic dropout, we obtained slightly different embeddings for the same description, which we used as a positive pair, while using other items in the batch as negative examples. This approach enabled the model to learn meaningful semantic representations without labeled data. However, due to the presence of similar descriptions in the dataset, some negative pairs were actually semantically close (false negatives), potentially reducing the effectiveness of the training signal. Despite the noise, the resulting embeddings still produced coherent local structures.

To mitigate the risk of semantically similar false negatives, we also implemented a supervised contrastive learning strategy using \texttt{InfoNCE} loss \cite{liao2021sentence} and a custom supervised contrastive learning dataset previously introduced in the data section of this paper. This setup allowed the model to explicitly align semantically related items and separate unrelated ones. While limited in scale by hardware constraints, this strategy offered a more reliable learning signal and reduced the likelihood of unintentionally penalizing similar examples.

Inspired by the work of \emph{Liao et al.} \cite{liao2021sentence}, we added a two-layer softmax classification head to the supervised model, a configuration they showed to yield strong performance on the \texttt{STS} task. This not only supported more stable training but also aligned the objective of organizing the dataset around fine-grained semantic relationships.

All fine-tuned models used in our experiments were trained for 8 to 16 epochs, depending on convergence behavior and early stopping criteria. Models using the unsupervised \texttt{SimCSE} strategy converged faster, aided by the larger batch size, while the supervised \texttt{InfoNCE} models had limited batch sizes due to \texttt{GPU} memory constraints. We monitored training and validation loss, as well as \texttt{STS-B} and \texttt{In-Context STS-B} Pearson scores (explained in the evaluation subsection) on held-out validation subsets. Each configuration was run multiple times to report mean and standard deviation. Only the best-performing hyperparameter combinations - namely learning rate and temperature - are included in the results.

\paragraph{Interpretability:} Unlike images, where visual differences and similarities are readily perceptible, comparing two textual descriptions is cognitively demanding. To address this, we compute attributions and highlight the specific words in each description that most influence its position in the embedding space (Figure \ref{fig:baskets}).

To compute word-level attributions, we adopt the Integrated Gradients (\texttt{IG}) method \cite{sundararajan2017axiomatic}, implemented via the \texttt{Captum} library \cite{kokhlikyan2020captum}. \texttt{IG} estimates each token’s contribution by integrating gradients along a path from a neutral baseline input to the actual description. We define our scalar output function as the cosine similarity between a given item’s final embedding and a reference point in the embedding space. This formulation allows us to attribute differences in semantic position back to specific tokens in the input, enabling more intuitive textual explanations. In practice, we discretize the integral, sum across token dimensions and compose token attributions to generate word-level scores.

\paragraph{Evaluation:} Since our objective was not classification but semantic representation, conventional metrics like accuracy or precision and recall are no longer applicable. Instead, we assess embedding quality through \texttt{STS}, which quantifies how well a model's embedding space reflects human judgments of sentence similarity. Since we are assessing models' semantic understanding, unlike the image-based pipeline, where no baseline classifier existed, the text-based setup allows us to quantitatively compare our fine-tuned models against their vanilla counterparts and clearly demonstrate performance gains.

Differently from the image-based pipeline, in this case, since we are directly measuring the quality of semantic understanding of the models, we can compare our fine-tuned models to their vanilla correspondant and quantitatively show the performance gains.

For standardized evaluation, we rely on the Brazilian Portuguese \texttt{STS-B} dataset from the \texttt{extraglue} benchmark \cite{osorio-etal-2024-portulan}, a translation of the well-known \texttt{GLUE} benchmark \cite{wang2018glue}. Each sentence pair in this dataset is annotated with a similarity score from 0 (unrelated) to 5 (equivalent). We compute cosine similarity between the sentence embeddings produced by our models and evaluate performance via Pearson correlation against the ground-truth scores. Although our models were not specifically fine-tuned on \texttt{STS-B}, they consistently outperformed their vanilla counterparts, indicating that our contrastive fine-tuning enhanced general semantic understanding. Following standard practice, due to the unavailability of labels for the official test set, we used validation set of \texttt{STS-B}, which we further split into validation and test subsets for our training process.

To better assess our models within domain-specific contexts, we also used a custom evaluation benchmark previously introduced in the data section: \texttt{In-Context STS-B}. Positive pairs were assigned a similarity score of 4 and negatives a score of 1, trying to reflect some conventions in \texttt{STS-B} scoring. We evaluated the models on this dataset using the same approach as before: computing cosine similarities between embeddings and measuring Pearson correlation with the assigned similarity scores. The dataset was also divided into validation and test splits. Notably, performance gains on this benchmark were often larger than on the standard \texttt{STS-B}, underscoring the value of this domain-adapted set as a more sensitive proxy for the kinds of semantic similarity our models aim to capture.

\paragraph{BERTimbau-based Models:} Results are summarized in Table \ref{tab:bertimbau-results}.

\begin{table}[h]
\centering
\scriptsize
\begin{tabular}{ c c c c }
\cline{1-4}
\textbf{Learning Method} & \textbf{Temperature} & \textbf{STS-B} & \textbf{In-Context STS-B} \\
\cline{1-4}
Vanilla & -- & 0.70 ± 0.01 & 0.75 ± 0.01 \\
Unsupervised & 0.05 & 0.71 ± 0.00 & \textbf{0.86 ± 0.01} \\
Unsupervised & 0.2 & 0.74 ± 0.00 & \textbf{0.86 ± 0.00} \\
Supervised & 0.05 & \underline{0.78 ± 0.01} & \textbf{0.86 ± 0.01} \\
Supervised & 0.2 & \textbf{0.79 ± 0.01} & \underline{0.82 ± 0.00} \\
\end{tabular}
\caption{\small Parameters and test results for \texttt{BERTimbau}-based models. Learning rate is always set to \(3 \times 10^{-6}\) (when applicable).}
\label{tab:bertimbau-results}
\end{table}

Evaluation results highlight several trends. All fine-tuned models outperformed their vanilla counterparts, confirming the effectiveness of contrastive training. Unsupervised models showed strong improvements on the domain-specific \texttt{In-Context STS-B}, but smaller gains on the general \texttt{STS-B}, indicating specialization at the cost of generalization. Higher temperatures helped these models by softening the impact of false negatives. In contrast, supervised models achieved a better balance, with some configurations matching the unsupervised setting in domain-specific performance while substantially improving general \texttt{STS-B} scores. For these models, lower temperatures yielded the best average performance, likely due to the clearer learning signal from curated positives and negatives. Overall, both methods were effective, with supervised contrastive learning offering more stable and generalizable improvements.

\paragraph{Albertina-based Models:} Results are summarized in Table \ref{tab:albertina-results}.

\begin{table}[h]
\centering
\scriptsize
\begin{tabular}{ c c c c }
\cline{1-4}
\textbf{Learning Method} & \textbf{Temperature} & \textbf{STS-B} & \textbf{In-Context STS-B} \\
\cline{1-4}
Vanilla & -- & 0.69 ± 0.01 & 0.68 ± 0.03 \\
Uns. SimCSE & 0.05 & 0.70 ± 0.01 & 0.85 ± 0.01 \\
Uns. SimCSE & 0.1 & \underline{0.72 ± 0.01} & \underline{0.86 ± 0.00} \\
InfoNCE & 0.05 & \textbf{0.74 ± 0.01} & \textbf{0.87 ± 0.01} \\
InfoNCE & 0.2 & \underline{0.72 ± 0.01} & 0.78 ± 0.01 \\
\end{tabular}
\caption{\small Parameters and test results for \texttt{Albertina}-based models. Learning rate is always set to \(1 \times 10^{-6}\) (when applicable).}
\label{tab:albertina-results}
\end{table}

As with the previous model, all fine-tuned \texttt{Albertina} variants outperformed the vanilla encoder on both benchmarks. However, overall performance was slightly lower compared to the \texttt{BERTimbau} models. While this may be surprising given that the larger \texttt{Albertina} variants typically outperform \texttt{BERTimbau} on \texttt{STS} benchmarks, the difference is more understandable at the base model scale. Once again, increasing the temperature in the unsupervised setup improved results, mitigating the impact of false negatives. Meanwhile, supervised models provided the most balanced performance across tasks, with lower temperature values yielding the highest scores on both general and domain-specific benchmarks.

\begin{figure}[h]
\centering
\includegraphics[width=0.22\textwidth]{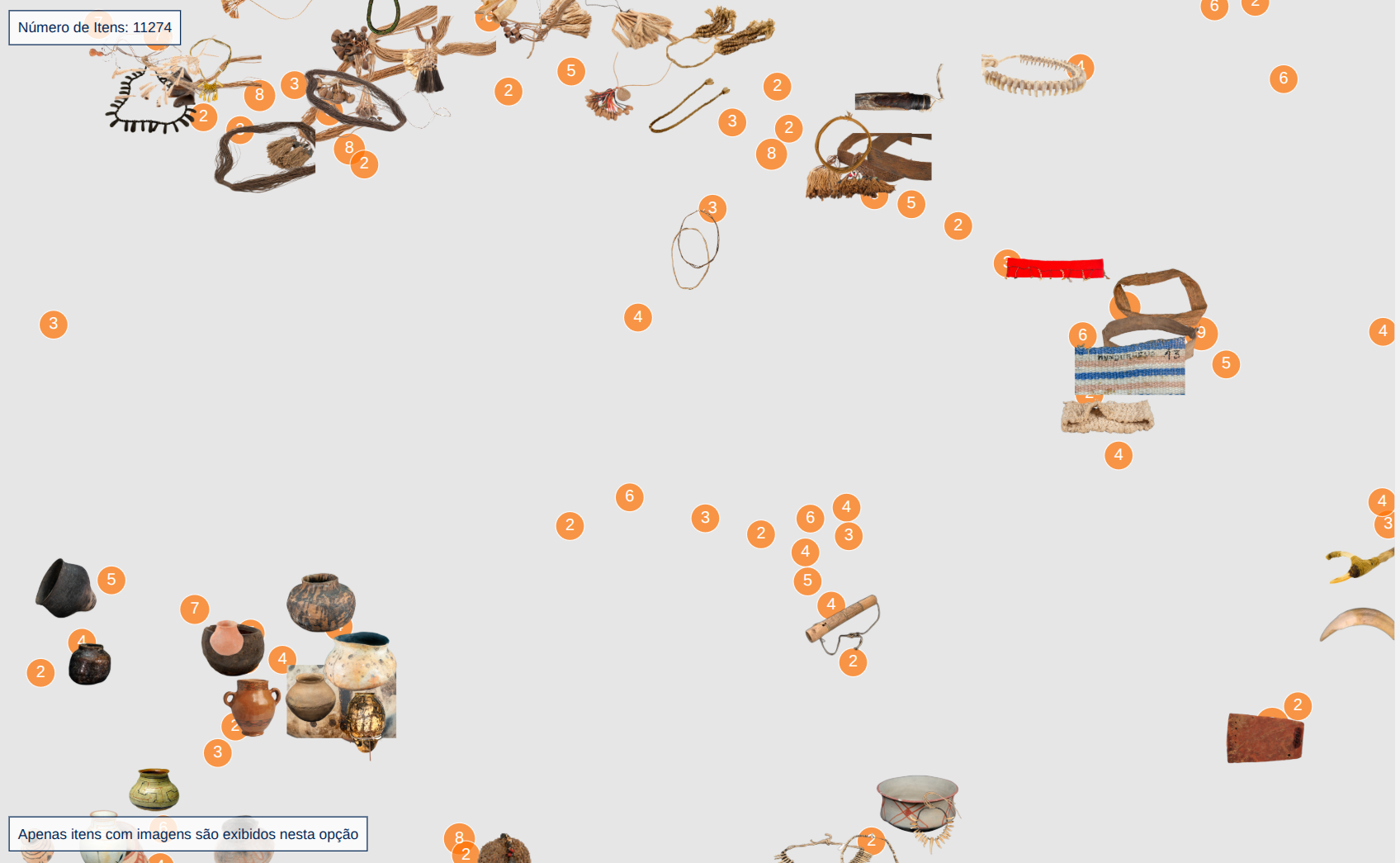}
\includegraphics[width=0.22\textwidth]{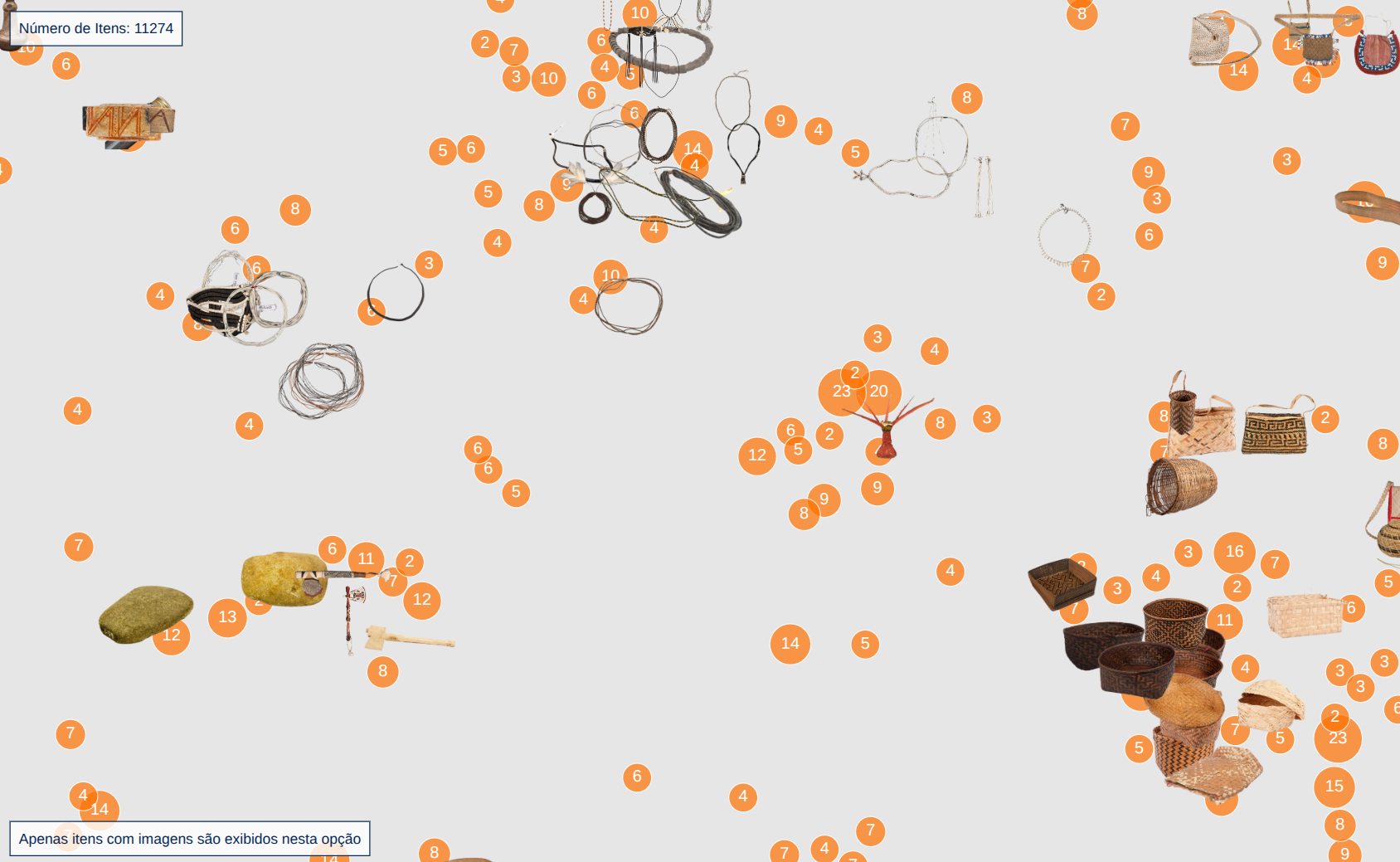}
\caption{\small Zoomed-in views of \texttt{UMAP} projections for \texttt{BERTimbau} trained with supervised contrastive learning. Both regions reveal coherent and interpretable semantic clusters.}
\label{fig:text-clusters}
\end{figure}

\begin{figure}[h]
\centering
\includegraphics[width=0.2\textwidth]{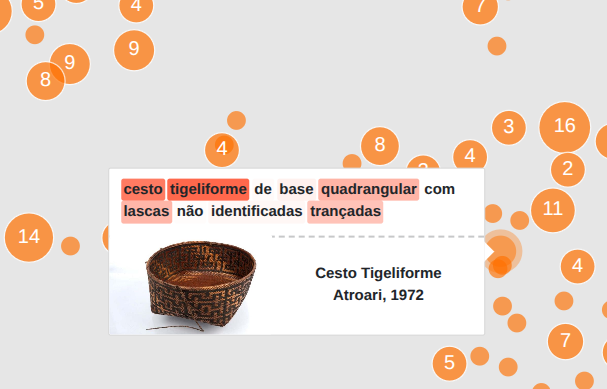}
\includegraphics[width=0.2\textwidth]{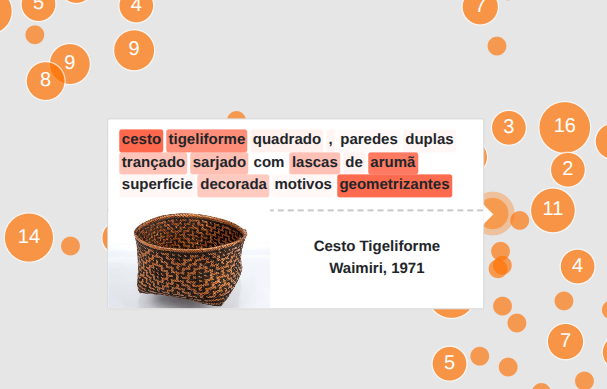}
\caption{\small Two items from the basket-related cluster of the right-hand image of Figure \ref{fig:text-clusters} with closely aligned semantic descriptions, both placed within the same cluster.}
\label{fig:baskets}
\end{figure}

\paragraph{Latent Observations:} Following the quantitative evaluations, we highlight several latent patterns uncovered by a \texttt{BERTimbau} model fine-tuned in the supervised contrastive setting with the lowest temperature. The projected embedding space revealed well-defined local clusters, even though global structure remained limited. Zoomed-in views of the projection show tight groupings of conceptually similar items, often despite notable differences in visual appearance. For example, one cluster in the left-hand image of Figure \ref{fig:text-clusters} contains vases made from diverse materials and styles, grouped together through shared descriptive vocabulary that emphasizes function or theme. Another cluster, shown in the right-hand image, brings together basket-related items whose descriptions align conceptually despite lexical differences. Figure \ref{fig:baskets} illustrates two such items: while their texts use different terms like “quadrangular” and “geometrizantes,” both evoke geometric structure, underscoring the model’s ability to capture deep semantic relationships.

\section{Visualization Tool}

\begin{figure}[h]
    \centering
    \includegraphics[width=0.22\textwidth]{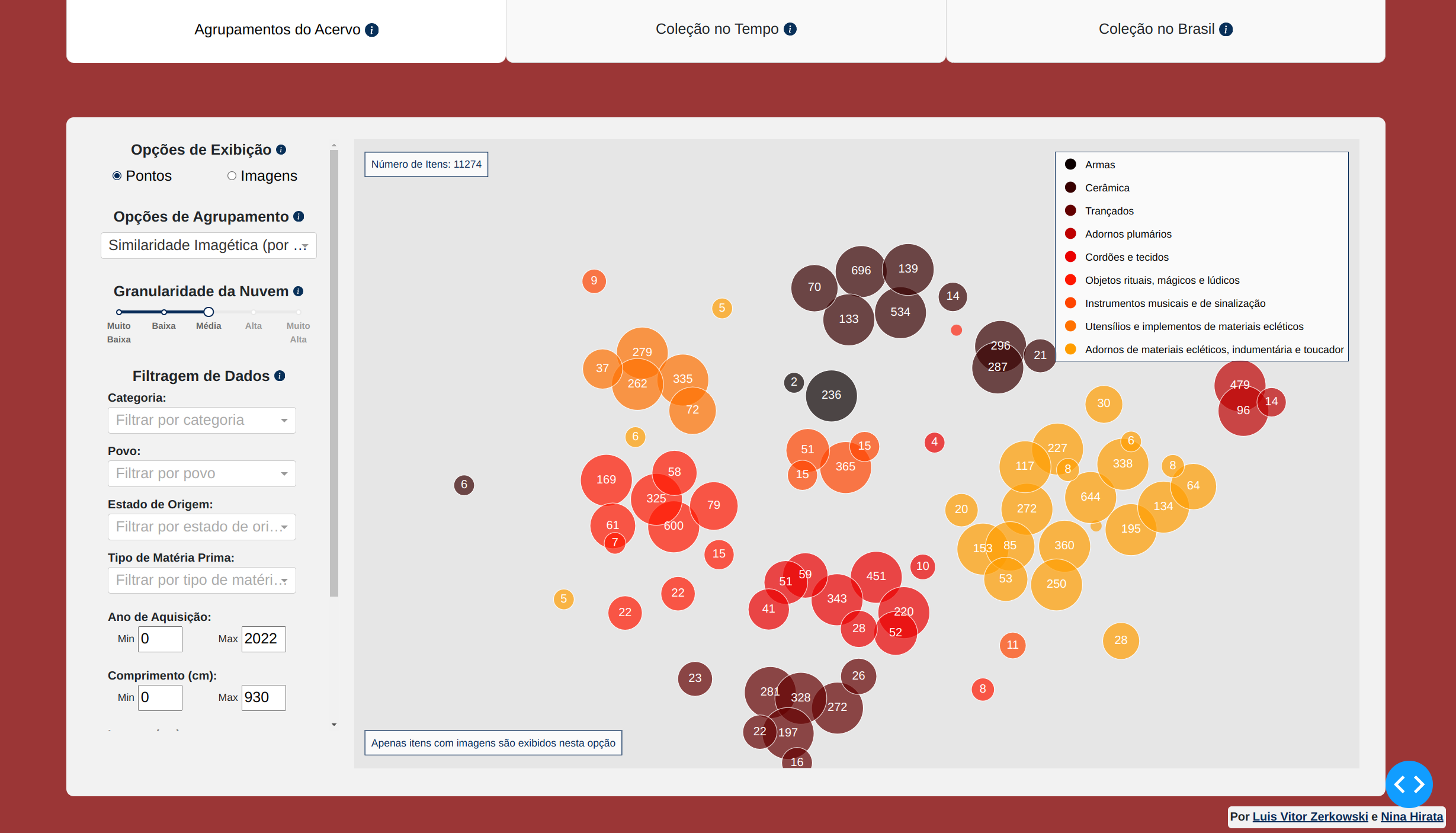}
    \includegraphics[width=0.22\textwidth]{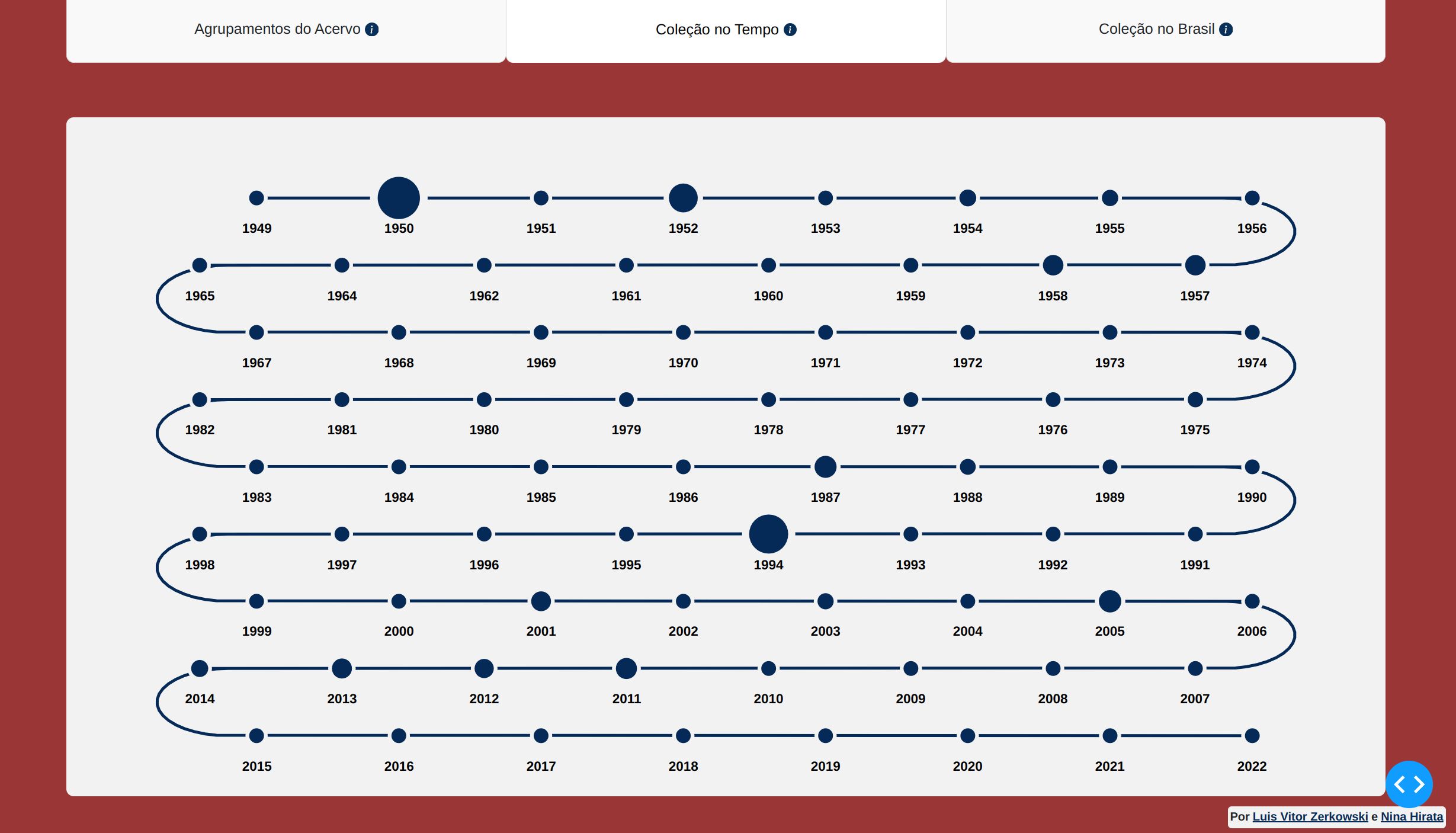}
    \includegraphics[width=0.22\textwidth]{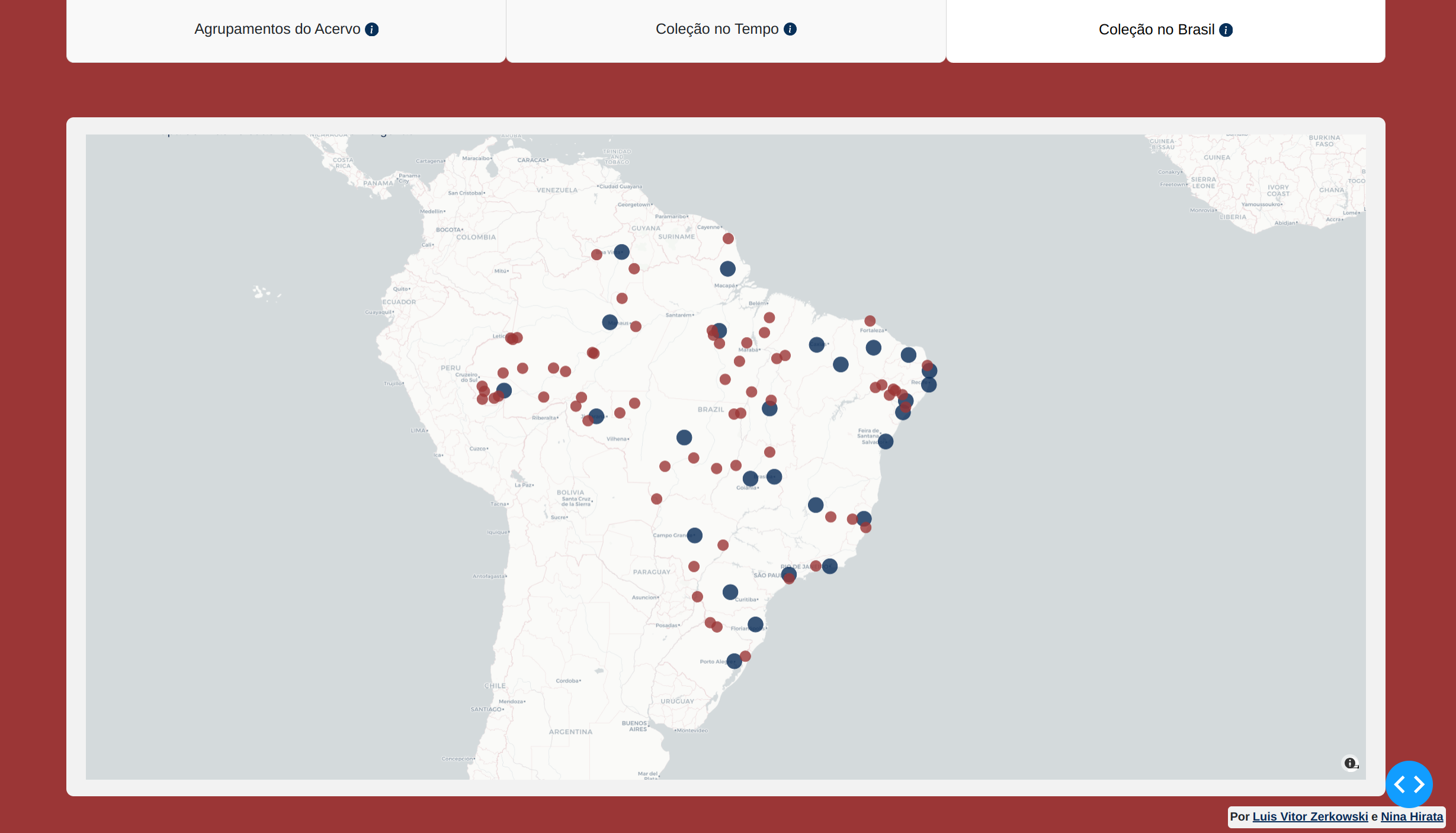}
    \caption{\small Primary interface tabs: collection semantic space (top-left), temporal view (top-right), geographic view (bottom).}
    \label{fig:visual-overview}
\end{figure}

Having described the technical groundwork, we conclude with the final component of our system: an interactive visualization tool designed to make our machine learning insights accessible and explorable. Developed in \texttt{Python} using \texttt{Plotly} and \texttt{Dash} \cite{plotly2015, dash2024}, the tool features three core sections: a semantic point-cloud view, a temporal acquisition timeline, and a geographic map of communities, each enabling different modes of exploration (Figure \ref{fig:visual-overview}). Users can navigate item similarity, filter by metadata, and explore spatial or historical patterns. Built in collaboration with the \emph{Museu Nacional dos Povos Indígenas} \cite{museudoindio2025}, the tool integrates feedback from multiple departments and adds features beyond the original \emph{Tainacan} platform \cite{tainacan2025}.

\subsection{Collection Semantic Spaces}

The core of the visualization tool is the semantic point-cloud interface, where each item is displayed as a 2D point using either marker-based or image-based views. A lazy loading system ensures responsiveness by only rendering visible points, while a granularity slider controls the density of displayed items. To prevent clutter, markers collapse into clusters when zoomed out, using a viewport-sensitive \texttt{KD-Tree} \cite{Bentley:1975:MBS:361002.361007} grouping strategy for more intuitive, screen-space aggregation. Additionally, items can be hovered (for images or summarized descriptions depending on the modality) or clicked to open full records on the \emph{Tainacan} platform.

Multiple semantic grouping modes allow exploration through different lenses. These include the previously introduced material composition, image-based visual similarity, and textual similarity. Each mode produces distinct clustering patterns suited to different exploration goals, whether conceptual, visual, or descriptive.

A flexible filtering panel enables users to refine the dataset by multiple attributes. Filters support both multiclass values and numeric ranges. A live count keeps track of the amount of remaining items after filtering. This combination of interactive controls and semantically rich embeddings makes the tool effective for both exploratory browsing and curatorial analysis.

\subsection{Temporal View}

The second main tab of the visualization tool presents a temporal interface for exploring the museum's acquisition history, allowing users to examine how the collection evolved over time. The view consists of two interactive components: a zig-zag timeline summarizing yearly acquisition counts and a detailed breakdown for each selected year.

The top-level zig-zag timeline displays a marker for each acquisition year, with marker size reflecting item count; hovering reveals item totals, while clicking transitions to a monthly breakdown. The detailed view shows a grid of item thumbnails color-coded by acquisition month and paired with a monthly histogram. Items lacking exact dates appear first and are grouped under a separate bar in the histogram. Rich interactions include: hovering over a thumbnail to enlarge it, highlight its month, and display metadata such as item name, date, community, collection, and donor, and hovering over a histogram bar to highlight all items from that month. Navigation is paginated for the thumbnail grid to ensure interface responsiveness and naturally the hover actions are scoped per page to also prevent rendering issues. Together, these views offer a visually engaging and analytically powerful tool for temporal exploration of the collection.

\subsection{Geographic View}

The third tab of the visualization tool offers a geographic view of the museum's collection, mapping items across Brazil to highlight Indigenous cultural presence. Red markers indicate specific communities (where data allows), while blue markers aggregate items by state. Clicking a marker opens a modal with item thumbnails, metadata, and optional full descriptions. Due to incomplete geographic mappings from the \emph{Instituto Socioambiental} (ISA) \cite{socioambiental2025} to our data, only around 40\% of communities could be precisely located. This spatial perspective complements the semantic and temporal views, enabling users to explore regional patterns and community representation.

\section{Conclusion and Future Work}

This project aimed to support the cultural preservation of Indigenous heritage in Brazil by enhancing the \emph{Museu Nacional dos Povos Indígenas} digital collection \cite{museudoindio2025} through modern machine learning. We developed visual and textual pipelines that generated structured, machine-learning-ready representations of the collection, made available via an open-source repository. These outputs enrich curatorial work, promote public engagement, and support Indigenous visibility in the digital space, not by replacing expertise, but by complementing it with computational tools.

Future work includes dynamic updates to the system as new items are added, expanded filtering options, and visual integration of cluster centroids for improved interpretability. More advanced features like semantic search, multimodal embeddings, immersive interfaces, and 3D reconstructions using techniques like Gaussian Splatting \cite{kerbl2023gaussiansplatting} could further enhance both research and public interaction. In all cases, our aim is to support intercultural dialogue and Indigenous protagonism through respectful and meaningful technological design.

\section{Acknowledgments}

We would like to express our sincere gratitude to all those who contributed to the development of this project. We are especially grateful to the \emph{Museu Nacional dos Povos Indígenas} and its team. In particular, Seiji for serving as our primary communication with the museum and helping the project’s deployment on the federal government’s website; Monique for her valuable input on improving the visual tool; and Luiza for authorizing the use of the collection’s images and contributing with insightful suggestions.

We would also like to thank São Paulo Research Foundation (FAPESP) -- grant \#2022/15304-4 for supporting this research.

\bibliography{aaai2026}

\makeatletter
\@ifundefined{isChecklistMainFile}{
  \newif\ifreproStandalone
  \reproStandalonetrue
}{
  \newif\ifreproStandalone
  \reproStandalonefalse
}
\makeatother

\ifreproStandalone
\documentclass[letterpaper]{article}
\usepackage[submission]{aaai2026}
\setlength{\pdfpagewidth}{8.5in}
\setlength{\pdfpageheight}{11in}
\usepackage{times}
\usepackage{helvet}
\usepackage{courier}
\usepackage{xcolor}
\frenchspacing

\begin{document}
\fi
\setlength{\leftmargini}{20pt}
\makeatletter\def\@listi{\leftmargin\leftmargini \topsep .5em \parsep .5em \itemsep .5em}
\def\@listii{\leftmargin\leftmarginii \labelwidth\leftmarginii \advance\labelwidth-\labelsep \topsep .4em \parsep .4em \itemsep .4em}
\def\@listiii{\leftmargin\leftmarginiii \labelwidth\leftmarginiii \advance\labelwidth-\labelsep \topsep .4em \parsep .4em \itemsep .4em}\makeatother

\setcounter{secnumdepth}{0}
\renewcommand\thesubsection{\arabic{subsection}}
\renewcommand\labelenumi{\thesubsection.\arabic{enumi}}

\newcounter{checksubsection}
\newcounter{checkitem}[checksubsection]

\newcommand{\checksubsection}[1]{%
  \refstepcounter{checksubsection}%
  \paragraph{\arabic{checksubsection}. #1}%
  \setcounter{checkitem}{0}%
}

\newcommand{\checkitem}{%
  \refstepcounter{checkitem}%
  \item[\arabic{checksubsection}.\arabic{checkitem}.]%
}
\newcommand{\question}[2]{\normalcolor\checkitem #1 #2 \color{blue}}
\newcommand{\ifyespoints}[1]{\makebox[0pt][l]{\hspace{-15pt}\normalcolor #1}}

\section*{Reproducibility Checklist}

\vspace{1em}
\hrule
\vspace{1em}


\checksubsection{General Paper Structure}
\begin{itemize}

\question{Includes a conceptual outline and/or pseudocode description of AI methods introduced}{(yes/partial/no/NA)}
yes

\question{Clearly delineates statements that are opinions, hypothesis, and speculation from objective facts and results}{(yes/no)}
yes

\question{Provides well-marked pedagogical references for less-familiar readers to gain background necessary to replicate the paper}{(yes/no)}
yes

\end{itemize}
\checksubsection{Theoretical Contributions}
\begin{itemize}

\question{Does this paper make theoretical contributions?}{(yes/no)}
no

	\ifyespoints{\vspace{1.2em}If yes, please address the following points:}
        \begin{itemize}
	
	\question{All assumptions and restrictions are stated clearly and formally}{(yes/partial/no)}
	Type your response here

	\question{All novel claims are stated formally (e.g., in theorem statements)}{(yes/partial/no)}
	Type your response here

	\question{Proofs of all novel claims are included}{(yes/partial/no)}
	Type your response here

	\question{Proof sketches or intuitions are given for complex and/or novel results}{(yes/partial/no)}
	Type your response here

	\question{Appropriate citations to theoretical tools used are given}{(yes/partial/no)}
	Type your response here

	\question{All theoretical claims are demonstrated empirically to hold}{(yes/partial/no/NA)}
	Type your response here

	\question{All experimental code used to eliminate or disprove claims is included}{(yes/no/NA)}
	Type your response here
	
	\end{itemize}
\end{itemize}

\checksubsection{Dataset Usage}
\begin{itemize}

\question{Does this paper rely on one or more datasets?}{(yes/no)}
yes

\ifyespoints{If yes, please address the following points:}
\begin{itemize}

	\question{A motivation is given for why the experiments are conducted on the selected datasets}{(yes/partial/no/NA)}
	yes

	\question{All novel datasets introduced in this paper are included in a data appendix}{(yes/partial/no/NA)}
	no

	\question{All novel datasets introduced in this paper will be made publicly available upon publication of the paper with a license that allows free usage for research purposes}{(yes/partial/no/NA)}
	partial

	\question{All datasets drawn from the existing literature (potentially including authors' own previously published work) are accompanied by appropriate citations}{(yes/no/NA)}
	yes

	\question{All datasets drawn from the existing literature (potentially including authors' own previously published work) are publicly available}{(yes/partial/no/NA)}
	yes

	\question{All datasets that are not publicly available are described in detail, with explanation why publicly available alternatives are not scientifically satisficing}{(yes/partial/no/NA)}
	yes

\end{itemize}
\end{itemize}

\checksubsection{Computational Experiments}
\begin{itemize}

\question{Does this paper include computational experiments?}{(yes/no)}
yes

\ifyespoints{If yes, please address the following points:}
\begin{itemize}

	\question{This paper states the number and range of values tried per (hyper-) parameter during development of the paper, along with the criterion used for selecting the final parameter setting}{(yes/partial/no/NA)}
	yes

	\question{Any code required for pre-processing data is included in the appendix}{(yes/partial/no)}
	no

	\question{All source code required for conducting and analyzing the experiments is included in a code appendix}{(yes/partial/no)}
	no

	\question{All source code required for conducting and analyzing the experiments will be made publicly available upon publication of the paper with a license that allows free usage for research purposes}{(yes/partial/no)}
	yes
        
	\question{All source code implementing new methods have comments detailing the implementation, with references to the paper where each step comes from}{(yes/partial/no)}
	yes

	\question{If an algorithm depends on randomness, then the method used for setting seeds is described in a way sufficient to allow replication of results}{(yes/partial/no/NA)}
	partial

	\question{This paper specifies the computing infrastructure used for running experiments (hardware and software), including GPU/CPU models; amount of memory; operating system; names and versions of relevant software libraries and frameworks}{(yes/partial/no)}
	partial

	\question{This paper formally describes evaluation metrics used and explains the motivation for choosing these metrics}{(yes/partial/no)}
	yes

	\question{This paper states the number of algorithm runs used to compute each reported result}{(yes/no)}
	yes

	\question{Analysis of experiments goes beyond single-dimensional summaries of performance (e.g., average; median) to include measures of variation, confidence, or other distributional information}{(yes/no)}
	yes

	\question{The significance of any improvement or decrease in performance is judged using appropriate statistical tests (e.g., Wilcoxon signed-rank)}{(yes/partial/no)}
	no

	\question{This paper lists all final (hyper-)parameters used for each model/algorithm in the paper’s experiments}{(yes/partial/no/NA)}
	yes

\end{itemize}
\end{itemize}
\ifreproStandalone
\end{document}
\fi

\end{document}